\documentclass[a4paper,11pt]{article}
\usepackage{jheppub} % for details on the use of the package, please see the JINST-author-manual
\usepackage{lineno}
\usepackage{subfig}

\title{\boldmath Analytical Structure and Spectroscopic Signatures of Quartic Nonlinear Electrodynamics in Holographic Fermion Systems
}

\author{Vitor Neves da Silva,  }
\author {Luiz Antonio Barreiro}
\affiliation{Institute of Geosciences and Exact Sciences,\\
Physics Department, São Paulo State University (UNESP),\\
 Rio Claro, São Paulo, Brazil}

\emailAdd{vitor-neves.silva@unesp.br }
\emailAdd{luiz.a.barreiro@unesp.br }

\abstract{
We investigate the effects of the leading string-inspired quartic
correction $\alpha'^2F^4$ on holographic fermions and charge transport
in a bottom-up AdS$_4$/CFT$_3$ model. Working in the probe limit, we
derive the nonlinear electrostatic background, obtain an exact first
integral of the gauge equation, establish the associated constitutive
relation, and identify the endpoint of the globally regular
electrostatic branch. A nonlinear response function naturally emerges
from the gauge sector and provides a unified description of the
background, gauge-field fluctuations, and transport coefficients.
Using these analytical backgrounds, we compute the fermionic retarded
Green's function and spectral function. The quartic interaction shifts
the Fermi momentum, enhances the spectral weight, increases the
renormalized Fermi velocity, and reduces both momentum and energy
linewidths, providing spectroscopic signatures consistent with enhanced 
quasiparticle coherence while preserving the holographic Fermi surface. We further investigate
optical and DC conductivities, showing that the nonlinear interaction
redistributes the optical response from the infrared toward
intermediate frequencies, whereas the DC conductivity is obtained
analytically as $\sigma_{\mathrm{DC}}=\mathcal K(1)=\rho/E(1)$ and
decreases monotonically with increasing nonlinear coupling. These
results establish a direct connection between nonlinear bulk
electrodynamics, momentum-resolved spectroscopic observables, and
transport properties in the dual strongly coupled theory.
}

\begin{document}
\maketitle
\flushbottom

\section{Introduction}\label{intro}

Since its original formulation, the gauge/gravity correspondence has become a useful framework for studying strongly coupled quantum systems through gravitational dynamics in asymptotically Anti-de Sitter spacetimes ~\cite{Maldacena1998,Gubser1998,Witten1998}. Beyond its role in quantum gravity, holography has been widely used to investigate strongly correlated matter, unconventional transport, holographic superconductivity, and non-Fermi-liquid behavior \cite{Hartnoll2009Lectures,Herzog2009Lectures,HartnollHerzogHorowitz2008}.

A standard application is the study of holographic fermions. Probe Dirac fields in charged AdS black-hole backgrounds make it possible to compute retarded fermionic Green's functions and to investigate holographic Fermi surfaces, non-Fermi liquids, Mott-like phases, and quantum critical behavior \cite{Lee2009,CubrovicZaanenSchalm2009,LiuMcGreevyVegh2011,FaulknerIqbalLiuMcGreevyVegh2010,FaulknerLiuMcGreevyVegh2011,EdalatiLeighPhillips2011}. The same framework has also been used to analyze transport governed by holographic Fermi surfaces and emergent infrared scaling regions \cite{FaulknerIqbalLiuMcGreevyVegh2013,KachruShyani2023}. In these constructions, the fermionic spectral function is determined by the background geometry and by the gauge field through which the fermions propagate. Consequently, modifications of the bulk gauge dynamics are expected to leave direct signatures in the boundary fermionic spectral response.

During the same period, angle-resolved photoemission spectroscopy (ARPES) has become the primary experimental probe of single-particle excitations in strongly correlated quantum materials. Modern ARPES experiments provide detailed information on spectral weight distributions, quasiparticle dispersions, scattering rates, pseudogap formation, and non-Fermi-liquid behavior in systems ranging from cuprate superconductors to strange metals. Since holographic fermionic Green's functions determine the single-particle spectral function, and ARPES provides experimental access to this quantity through photoemission intensity, momentum-resolved spectroscopy constitutes one of the most direct experimental points of contact for holographic fermionic models.

Considerable effort has therefore been devoted to identifying which ingredients of holographic models control experimentally relevant spectral observables. Most previous investigations focus on changes in the gravitational background, fermionic couplings, dipole interactions, or infrared geometry. By contrast, comparatively less attention has been devoted to understanding how nonlinear corrections in the bulk gauge sector modify the fermionic spectral response.

Most holographic fermion studies adopt Maxwell theory for the bulk gauge sector, even though nonlinear electromagnetic interactions arise naturally in effective descriptions of open-string and D-brane dynamics \cite{BornInfeld1934,FradkinTseytlin1985,Tseytlin1997}. Such corrections encode finite-coupling effects in the dual field theory and modify the electrostatic background, potentially leaving observable imprints on boundary fermionic response \cite{GibbonsRasheed1995,JingChen2010,JingPanChen2011}.

Related nonlinear gauge sectors, such as Born--Infeld, power-Maxwell, and logarithmic electrodynamics, have been studied extensively in holography, especially in connection with superconducting phases and transport observables ~\cite{JingChen2010,JingPanChen2011,JingChenPan2011,Sheykhi2017Logarithmic,Roussev2024}. These studies have demonstrated that nonlinear electrodynamics can significantly modify the thermodynamic and transport properties of strongly coupled holographic systems. Unlike Born--Infeld electrodynamics, whose nonlinear response is governed by a non-polynomial structure, the present quartic theory corresponds to the leading operator in the derivative expansion of the open-string effective action. This analytical structure makes it possible to derive an exact first integral, obtain a closed nonlinear constitutive relation, identify analytically the endpoint of the globally regular electrostatic branch, and directly connect these analytical results with fermionic spectroscopy and transport observables. In this sense, the quartic model provides an analytically transparent setting in which one can determine which features of the fermionic response and of charge transport already follow from the leading nonlinear correction in the gauge sector.

Motivated by these considerations, in the present work we investigate the leading quartic correction proportional to $F^4$. Our primary objective is to understand how nonlinear bulk gauge dynamics modifies the fermionic spectral response, with particular emphasis on experimentally relevant momentum-resolved observables. In particular, we investigate how the nonlinear gauge sector affects quantities such as the Fermi momentum and the spectral weight. Our analysis is not intended as a complete ultraviolet completion of the bulk theory, but rather as a minimal effective framework in which the imprint of the leading quartic interaction on fermionic observables can be disentangled from more general higher-order effects.

We consider a bottom-up holographic model in asymptotically $AdS_{4}$ spacetime. The bulk gauge sector is deformed by the leading string-inspired quartic correction $\alpha'^{2}F^{4}$, while the geometry is kept fixed in the probe approximation. This choice should be understood as a controlled first step rather than as a complete finite-coupling treatment of the bulk theory. Within this framework, the probe approximation allows us to isolate the specific imprint of the nonlinear gauge dynamics on the electrostatic background, the fermionic spectral response, and the transport coefficients, without mixing these effects with gravitational backreaction. Although a fully backreacted treatment may quantitatively modify the location of spectral features and the values of the transport coefficients, the exact constitutive structure and the associated nonlinear response function derived here already provide a clean framework for identifying how the leading quartic interaction is encoded in boundary observables. A fully backreacted analysis would be valuable for assessing the quantitative robustness of these predictions, but lies beyond the scope of the present work.

We first analyze the nonlinear electrostatic background analytically and subsequently compute the fermionic Green's function in the resulting geometries. This combined analytical and numerical approach allows us to establish how quartic nonlinear electrodynamics modifies the fermionic spectral response and its associated momentum-resolved observables. Particular attention is devoted to the evolution of the Fermi momentum, spectral weight, and other characteristics of the fermionic spectrum.

The paper is organized as follows. In Sec.~\ref{model} we introduce the holographic model. In Sec.~\ref{sec:analytical_gauge} we analyze the nonlinear gauge sector and derive the relevant analytical results. In Sec.~\ref{sec:fermionic_response} we study the fermionic response in the nonlinear background. Section~\ref{sec:spectroscopy_transport} discusses the spectroscopic and transport signatures of the holographic fermionic response, emphasizing their connections with momentum-resolved photoemission spectroscopy (ARPES) and optical conductivity. Section~\ref{sec:conclusions} contains our conclusions.

\section{Holographic Model} \label{model}

We consider a bottom-up holographic model in asymptotically $AdS_{4}$ spacetime, in which a conserved $U(1)$ current in the $(2+1)$-dimensional boundary theory is described by a bulk gauge field. The geometry is kept fixed, while nonlinear corrections are incorporated in the gauge sector within the probe approximation, following the standard logic of probe matter sectors in holographic condensed-matter applications \cite{Hartnoll2009Lectures,Herzog2009Lectures,HartnollHerzogHorowitz2008}. Consequently, the higher-derivative electromagnetic interaction modifies the bulk gauge-field profile but does not backreact on the metric.

\subsection{Charged black-hole background}
The background geometry is chosen to be the electrically charged Reissner--Nordström
black hole in $AdS_{4}$. This charged black-hole geometry is the standard finite-density background used in many holographic studies of transport and fermionic response \cite{HartnollKovtun2007,Lee2009,LiuMcGreevyVegh2011} and is written as 
\begin{equation}
ds^{2}=\frac{\beta^{2}}{z^{2}}\left[-f(z)dt^{2}+dx^{2}+dy^{2}\right]+\frac{dz^{2}}{z^{2}f(z)}.\label{eq:metric_RNAdS}
\end{equation}
The radial coordinate $z$ is such that the conformal boundary is
located at $z=0$, while the event horizon is fixed at 
\begin{equation}
z_{H}=1.
\end{equation}
In the purely electric sector, the blackening function is 
\begin{equation}
f(z)=1+q^{2}z^{4}-(1+q^{2})z^{3},\label{eq:blackening_function}
\end{equation}
where $q$ denotes the electric charge parameter of the black hole.
The corresponding Hawking temperature is 
\begin{equation}
T=\frac{\beta|f'(1)|}{4\pi}=\frac{\beta(3-q^{2})}{4\pi}.\label{eq:hawking_temperature}
\end{equation}
Therefore, the nonextremal regime is defined by 
\begin{equation}
q^{2}<3.
\end{equation}
The parameter $\beta$ sets the overall energy scale of the dual field
theory and may be used to express all dimensionful quantities in units
adapted to the boundary coordinates.

\subsection{Nonlinear gauge sector}

The bulk dynamics is described by the Einstein--Hilbert term, the
Maxwell term, and the leading quartic correction to the electromagnetic
field strength, 
\begin{equation}
\mathcal{L}_{\mathrm{eff}}=\mathcal{L}_{\mathrm{EH}}+\mathcal{L}_{\mathrm{M}}+\mathcal{L}_{\mathrm{P}}.\label{eq:effective_lagrangian}
\end{equation}
The Maxwell contribution is 
\begin{equation}
\mathcal{L}_{\mathrm{M}}=-\frac{1}{4}F_{\mu\nu}F^{\mu\nu},\label{eq:maxwell_lagrangian}
\end{equation}
where 
\begin{equation}
F_{\mu\nu}=\nabla_{\mu}A_{\nu}-\nabla_{\nu}A_{\mu}.\label{eq:field_strength}
\end{equation}
The leading nonlinear correction is taken to be 
\begin{equation}
\mathcal{L}_{\mathrm{P}}=\frac{\alpha'^{2}}{8}\left[F_{\mu\lambda}F_{\nu}{}^{\lambda}F_{\rho}{}^{\mu}F^{\nu\rho}-\frac{1}{4}\left(F_{\mu\nu}F^{\mu\nu}\right)^{2}\right].\label{eq:quartic_lagrangian}
\end{equation}
This form is the Abelian quartic sector inspired by the low-energy open-string effective action \cite{FradkinTseytlin1985,Tseytlin1997,Barreiro2012}.
Here $\alpha'$ controls the strength of the higher-derivative correction.
Since the nonlinear gauge sector is treated in the probe approximation,
$\alpha'$ modifies the gauge-field equation but not the metric \eqref{eq:metric_RNAdS}.

Accordingly, the quartic interaction should be interpreted here as the leading correction in a truncated effective description of the gauge sector. The purpose of this truncation is not to claim a complete ultraviolet completion, but to determine which features of the fermionic response already follow from the leading nonlinear modification of the electrostatic background. This interpretation will also be important when discussing the critical coupling below, since that threshold should be associated with the loss of regularity of the present truncated branch of solutions.

Varying the action with respect to $A_{\nu}$ gives the nonlinear
Maxwell equation 
\begin{equation}
\nabla_{\mu}\left[F^{\mu\nu}-\alpha'^{2}\left(F^{\mu\lambda}F_{\lambda\rho}F^{\rho\nu}-\frac{1}{4}F^{\mu\nu}F_{\alpha\beta}F^{\alpha\beta}\right)\right]=0.\label{eq:nonlinear_maxwell_general}
\end{equation}

To describe both the equilibrium finite-density state and its linear
response within a unified framework, we decompose the bulk gauge field
into an equilibrium background and a small time-dependent perturbation,
\begin{equation}
A_{\mu}
=
A_{\mu}^{(0)}
+
\lambda\,\delta A_{\mu},
\label{eq:gauge_decomposition}
\end{equation}
where
\begin{equation}
A_{\mu}^{(0)}
=
(\phi(z),0,0,0),
\qquad
\delta A_{\mu}
=
\left(
0,
a_x(z)e^{-i\omega t},
0,
0
\right),
\label{eq:gauge_ansatz}
\end{equation}
and $\lambda\ll1$ is a bookkeeping parameter controlling the
linear-response expansion.

The background gauge field $A_\mu^{(0)}$ describes a static,
homogeneous, and isotropic finite-density state. Its electrostatic
potential $\phi(z)$ depends only on the holographic radial coordinate
and generates the radial electric field
\begin{equation}
E(z)=-\phi'(z),\label{eq:defE}
\end{equation}
associated with the bulk field-strength component $F_{zt}$.

Since $\lambda\ll1$, the perturbation $\delta A_\mu$ does not modify
the equilibrium background and can therefore be treated consistently
within linear response. The associated field-strength component
$F_{tx}$ generates a weak oscillating electric field that acts as an
external probe of the boundary theory. As will be discussed in
Sec.~V, this external perturbation provides the source that induces
the boundary current, allowing the optical conductivity of the dual field theory 
to be computed holographically..

This decomposition naturally separates the construction of the
equilibrium holographic background from the calculation of its linear
response. It also provides a natural bridge between the holographic description
and experimentally measurable spectroscopic and transport observables,
thereby establishing the framework used throughout the remainder of
this work.

Keeping only terms linear in the perturbation parameter $\lambda$, all contributions of order $\mathcal O(\lambda^2)$ and higher are neglected.
Substituting the ansatz (\ref{eq:gauge_ansatz}) into Eq.~(\ref{eq:nonlinear_maxwell_general}) and expanding consistently to first order in $\lambda$, one finds that the background electrostatic potential and the transverse fluctuation satisfy
\begin{equation}
\frac{d}{dz}\left[\phi'\mathcal{K}\right]=0,\label{eq:phi_nonlinear_reduced}
\end{equation}
and
\begin{equation}
\frac{d}{dz}\left[f\mathcal{K}a_{x}'\right]+\frac{\omega^{2}}{\beta^{2}f}\mathcal{K}a_{x}=0
\label{eq:A_nonlinear_reduced}
\end{equation}
where we have defined the effective nonlinear response function
\begin{equation}
\mathcal{K}(z)=1-\gamma z^{4}\phi'^2,
\label{nonlinearResponse}
\end{equation}
which encodes the quartic correction to the Maxwell sector. Here
\begin{equation}
\gamma\equiv
\frac{3\alpha'^2}{2\beta^2}.
\label{eq:gamma_definition}
\end{equation}
Notice that Eq.~(\ref{eq:phi_nonlinear_reduced}) is equivalent to
\begin{equation}
\frac{d}{dz}
\left(
\phi'
-\gamma z^4\phi'^3
\right)=0,
\label{eq:phi_nonlinear_expanded}
\end{equation}
which will be integrated analytically in Sec.~\ref{sec:analytical_gauge}.

The electrostatic potential satisfies the regularity condition 
\begin{equation}
\phi(1)=0\label{eq:horizon_phi_condition}
\end{equation}
at the horizon, while near the asymptotic boundary it behaves as 
\begin{equation}
\phi(z)=\mu-\rho z+\cdots.\label{eq:phi_boundary_expansion}
\end{equation}
According to the holographic dictionary, $\mu$ and $\rho$ are interpreted
as the chemical potential and charge density of the boundary theory,
respectively.

Although Eq.~(\ref{eq:phi_nonlinear_reduced}) is a second-order
nonlinear differential equation, its structure is considerably richer
than that of the linear Maxwell theory. An important feature of Eq.~(\ref{eq:phi_nonlinear_reduced}) is that it depends only on derivatives of the electrostatic potential. Consequently, it admits an exact first integral associated with a conserved radial quantity. As will be shown in Sec.~\ref{sec:analytical_gauge}, this conserved quantity naturally acquires the interpretation of an effective electric displacement field, leading to a nonlinear constitutive relation for the holographic bulk. This observation provides a transparent physical interpretation of the higher-derivative gauge interaction and establishes the analytical framework employed throughout the subsequent analysis of both the electromagnetic background and the fermionic sector.

\subsection{Probe fermionic sector}
The nonlinear electrostatic background determined by Eq.~\eqref{eq:phi_nonlinear_reduced}
is used as the external gauge field for a charged probe Dirac fermion.
In the probe-fermion approximation, the fermion does not backreact
on either the geometry or the gauge field. The fermionic Lagrangian
is 
\begin{equation}
\mathcal{L}_{F}=\bar{\psi}\left(i\Gamma^{\underline{a}}e_{\underline{a}}{}^{\mu}D_{\mu}-m\right)\psi,\label{eq:fermion_lagrangian}
\end{equation}
where underlined indices denote tangent-space indices. The covariant
derivative is 
\begin{equation}
D_{\mu}=\partial_{\mu}+\frac{1}{4}\omega_{\underline{a}\underline{b}\,\mu}\Gamma^{\underline{a}\underline{b}}-iq_{f}A_{\mu},\qquad\bar{\psi}=\psi^{\dagger}\Gamma^{\underline{t}}.\label{eq:fermion_covariant_derivative}
\end{equation}
Here $q_{f}$ is the fermion charge, introduced separately from the
black-hole charge parameter $q$. The Dirac equation is therefore
\begin{equation}
\left(i\Gamma^{\underline{a}}e_{\underline{a}}{}^{\mu}D_{\mu}-m\right)\psi=0.\label{eq:dirac_equation_general}
\end{equation}

Using translational invariance along the boundary directions, we write
\begin{equation}
\psi(z,t,x,y)=(-gg^{zz})^{-1/4}e^{-i\omega t+ikx}\tilde{\psi}(z),\label{eq:spinor_ansatz}
\end{equation}
where the prefactor $(-gg^{zz})^{-1/4}$ removes the spin-connection
contribution from the radial equation. We choose the gamma-matrix
basis 
\begin{equation}
\Gamma^{\underline{z}}=\sigma_{3}\otimes I_{2},\qquad\Gamma^{\underline{t}}=i\sigma_{1}\otimes I_{2},\qquad\Gamma^{\underline{x}}=\sigma_{2}\otimes\sigma_{3}.\label{eq:gamma_basis}
\end{equation}
The four-component spinor can then be decomposed as 
\begin{equation}
\tilde{\psi}(z)=\begin{pmatrix}\eta(z)\\[2mm]
\zeta(z)
\end{pmatrix},\quad\eta(z)=\begin{pmatrix}\eta_{1}(z)\\
\eta_{2}(z)
\end{pmatrix},\quad\zeta(z)=\begin{pmatrix}\zeta_{1}(z)\\
\zeta_{2}(z)
\end{pmatrix}.\label{eq:spinor_decomposition}
\end{equation}
Introducing the ratios 
\begin{equation}
\xi_{I}(z)\equiv\frac{\eta_{I}(z)}{\zeta_{I}(z)},\qquad I=1,2,\label{eq:xi_definition}
\end{equation}
the radial Dirac equation can be written in Riccati form as 
\begin{equation}
\begin{aligned}\xi_{I}'={} & \frac{2m}{z\sqrt{f}}\,\xi_{I}-\left(\frac{\omega+q_{f}\phi}{\beta f}-\frac{(-1)^{I}k}{\beta\sqrt{f}}\right)\\
 & -\left(\frac{\omega+q_{f}\phi}{\beta f}+\frac{(-1)^{I}k}{\beta\sqrt{f}}\right)\xi^{2}_{I}.
\end{aligned}
\label{eq:riccati_flow_reduced}
\end{equation}
In contrast to current-carrying backgrounds, no effective momentum
shift appears because $A_{x}=0$.

Near the asymptotic boundary, the solutions behave as 
\begin{equation}
\tilde{\psi}_{I}(z)=a_{I}z^{-m}\begin{pmatrix}0\\
1
\end{pmatrix}+b_{I}z^{m}\begin{pmatrix}1\\
0
\end{pmatrix}+\cdots,\qquad I=1,2.\label{eq:spinor_boundary_expansion}
\end{equation}
In the standard quantization, $a_{I}$ are identified with sources
and $b_{I}$ with expectation values of the dual fermionic operators.
The retarded Green's function is therefore extracted from 
\begin{equation}
G_{R}(\omega,k)=\lim_{z\to0}z^{-2m}\begin{pmatrix}\xi_{1}(z) & 0\\
0 & \xi_{2}(z)
\end{pmatrix}.\label{eq:retarded_green_function}
\end{equation}

We follow the standard real-time holographic prescription for spinorial correlators. At the horizon, the retarded correlator is obtained by imposing infalling boundary conditions, which implement the retarded prescription in the Lorentzian black-hole geometry \cite{SonStarinets2002,IqbalLiu2009Spinors}. 
For nonzero frequency this condition gives 
\begin{equation}
\xi_{I}(1)=i,\qquad I=1,2,\qquad(\omega\neq0).\label{eq:horizon_infalling_condition}
\end{equation}
The zero-frequency limit is obtained by smooth continuation from the
infalling solution. Solving Eq.~\eqref{eq:riccati_flow_reduced}
with these UV and IR boundary conditions determines the retarded Green's
function and the corresponding fermionic spectral function.

\section{Analytical Structure of the Nonlinear Gauge Sector}\label{sec:analytical_gauge}
Before solving the nonlinear gauge equations numerically, it is useful
to extract the main analytical properties of the electrostatic background.
Besides providing a consistency check for the numerical integration,
the analytical treatment reveals that the higher-derivative interaction
admits an exact conserved quantity, allowing the nonlinear Maxwell
equation to be reduced to an algebraic constitutive relation for the
radial electric field. This structure clarifies how the nonlinear
gauge dynamics modifies the electrostatic background that subsequently
enters the fermionic flow equations. This interpretation is analogous to the displacement-field formulation commonly used in nonlinear electrodynamics \cite{BornInfeld1934,Plebanski1970,GibbonsRasheed1995}.

\subsection{Exact First Integral}

Equation~(\ref{eq:phi_nonlinear_reduced}) or your expanded version, Eq.~(\ref{eq:phi_nonlinear_expanded}),  possesses a particularly
simple structure since it depends only on derivatives of the electrostatic
potential. It can be  immediately integrated, yielding the first integral 
\begin{equation}
\phi'-\gamma z^{4}\phi'^{3}=C,\label{eq:first_integral_phi}
\end{equation}
where $C$ is independent of the holographic coordinate.

To determine the integration constant, we use the asymptotic expansion
(\ref{eq:phi_boundary_expansion}). Near the conformal boundary, 
\begin{equation}
\phi'(z)=-\rho+\mathcal{O}(z),
\end{equation}
whereas the nonlinear contribution is suppressed by the explicit factor
$z^{4}$. Consequently, 
\begin{equation}
C=-\rho,
\end{equation}
and the exact first integral assumes the form 
\begin{equation}
\phi'-\gamma z^{4}\phi'^{3}=-\rho.\label{eq:first_integral_final}
\end{equation}

Using Eq.~(\ref{eq:defE}) for the radial electric field, Eq.~(\ref{eq:first_integral_final}) becomes 
\begin{equation}
E(z)-\gamma z^{4}E(z)^{3}=\rho.\label{eq:constitutive}
\end{equation}
In this form, the conserved charge density plays the role of a radial electric displacement field, while 
the physical electric field is determined by a nonlinear constitutive equation \cite{Plebanski1970,GibbonsRasheed1995}.

\begin{figure}
\begin{centering}
\includegraphics[width=0.7\columnwidth]{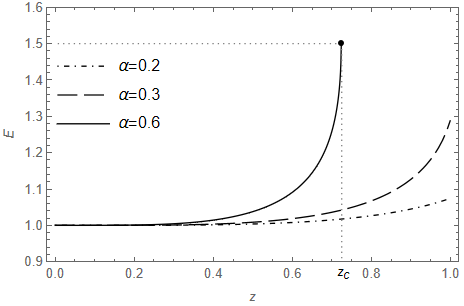}
\par\end{centering}
\caption{Radial electric field $E(z)$ for different values of the nonlinear
coupling $\alpha'$, with $\rho=\beta=1$. For sufficiently small
nonlinear coupling, the electric field remains regular throughout
the holographic interval. As the coupling increases, the physical
branch develops a critical point at $z=z_{c}$,
where the differential constitutive response vanishes. Beyond this
point, no globally continuous positive solution connected to the Maxwell
branch exists.}\label{fig:Radial-electric-field}
\end{figure}

Equation~(\ref{eq:constitutive}) is the central analytical result
of the nonlinear gauge sector. Instead of solving the original second-order
differential equation, the electrostatic background is completely
determined by an algebraic relation between the radial electric field
and the conserved charge density. The corresponding radial electric
field is shown in Fig. \ref{fig:Radial-electric-field} for representative
values of the nonlinear coupling. 
For weak coupling, the electric field remains smooth over the entire
holographic interval and differs only slightly from the Maxwell result.
As the nonlinear interaction is strengthened, however, the electric
field grows more rapidly toward the infrared region, and the physical
branch approaches a critical configuration where its slope becomes
singular. This qualitative change signals the onset of a breakdown
of the monotonic response of the constitutive equation. 

The origin of this critical behavior is not immediately apparent from
Eq.~(\ref{eq:constitutive}) itself, but becomes transparent when
the equation is interpreted as a nonlinear constitutive relation between
the conserved electric displacement and the electric field. This interpretation
provides a simple physical picture of the higher-order electromagnetic
interaction and naturally leads to the analytical determination of
the critical point discussed in the next subsection.
\subsection{Nonlinear Constitutive Relation}

Equation~(\ref{eq:constitutive}) naturally admits the interpretation
of a constitutive equation describing an effective nonlinear
electromagnetic medium. Introducing the radial electric displacement,
\begin{equation}
D(E,z)\equiv E-\gamma z^{4}E^{3},
\label{eq:defD}
\end{equation}
the exact first integral assumes the compact form
\begin{equation}
D=\rho.
\label{eq:Dconst}
\end{equation}
Such a displacement-field formulation is standard in nonlinear electrodynamics and provides a useful way to diagnose branch structure and loss of invertibility \cite{Plebanski1970,Novello2000}.

Unlike Maxwell electrodynamics, where the electric displacement and
the electric field coincide, the higher-derivative interaction
introduces a nonlinear relation between these quantities. The
conserved charge density is therefore identified with a conserved
radial electric displacement, while the electric field becomes an
implicit function of both the conserved flux and the holographic
coordinate.

The response function introduced in Eq.~(\ref{nonlinearResponse}) admits a direct constitutive interpretation as an 
effective dielectric function,
\begin{equation}
\varepsilon_{\mathrm{eff}}(E,z)\equiv
\frac{D}{E}
=
1-\gamma z^{4}E^{2}=\mathcal{K}(z),
\label{eq:eps_eff}
\end{equation}
which reduces continuously to unity in the Maxwell limit,
$\gamma\rightarrow0$. Since the nonlinear contribution is proportional
to $z^{4}$, deviations from linear electrodynamics become progressively
stronger toward the infrared region of the bulk while remaining
negligible close to the AdS boundary. From this perspective, the holographic bulk 
behaves as an effective nonlinear dielectric medium, whose response departs increasingly 
from Maxwell electrodynamics toward the infrared region.

Even more relevant is the differential constitutive response,
\begin{equation}
\varepsilon_{\mathrm{diff}}(E,z)
\equiv
\frac{\partial D}{\partial E}
=
1-3\gamma z^{4}E^{2},
\label{eq:eps_diff}
\end{equation}
which controls the local invertibility of the constitutive relation.
As long as
\begin{equation}
\varepsilon_{\mathrm{diff}}(E,z)\neq0,
\label{eq:invertibility}
\end{equation}
the implicit function theorem guarantees the existence of a unique
local solution $E(\rho,z)$ continuously connected to the Maxwell
branch. The vanishing of the differential response is the local signal that the constitutive 
map ceases to be invertible, a familiar feature of nonlinear electromagnetic media \cite{Plebanski1970,GibbonsRasheed1995,Novello2000}.

The physical significance of this quantity becomes evident after
differentiating Eq.~(\ref{eq:constitutive}) with respect to the
holographic coordinate. Since the charge density is conserved, one
obtains
\begin{equation}
E'=
\frac{4\gamma z^{3}E^{3}}
{1-3\gamma z^{4}E^{2}}
=
\frac{4\gamma z^{3}E^{3}}
{\varepsilon_{\mathrm{diff}}},
\label{eq:Eprime}
\end{equation}
showing that the radial derivative of the electric field diverges
whenever the differential response vanishes. The critical point is
therefore determined exactly by the condition
\begin{equation}
1-3\gamma z_c^{4}E_c^{2}=0,
\label{eq:critical1}
\end{equation}
or equivalently,
\begin{equation}
\gamma z_c^{4}E_c^{2}
=
\frac13.
\label{eq:critical3}
\end{equation}

Substituting Eq.~(\ref{eq:critical3}) into the constitutive relation
(\ref{eq:constitutive}) immediately yields
\begin{equation}
E_c=\frac32\rho,
\label{eq:Ec}
\end{equation}
demonstrating that the critical electric field depends exclusively on
the conserved charge density and is completely independent of the
nonlinear coupling.

Substituting Eq.~(\ref{eq:Ec}) into
Eq.~(\ref{eq:critical3}) gives
\begin{equation}
z_c^{4}
=
\frac{4}{27\gamma\rho^{2}},
\label{eq:zc4}
\end{equation}
or, using Eq.~(\ref{eq:gamma_definition}),
\begin{equation}
z_c=
\left(
\frac{8\beta^{2}}
{81\alpha'^2\rho^{2}}
\right)^{1/4}.
\label{eq:zc}
\end{equation}

For the representative parameters employed in
Fig.~\ref{fig:Radial-electric-field},
namely $\alpha'=0.6$ and $\rho=\beta=1$, the analytical prediction is
\[
E_c=\frac32,
\qquad
z_c\simeq0.724,
\]
in excellent agreement with the numerical solution shown in
Fig.~\ref{fig:Radial-electric-field}. Beyond this point the physical
branch develops a vertical tangent and can no longer be continued
smoothly toward the black-hole horizon.

\begin{figure}[h!]
\begin{centering}
\includegraphics[width=0.7\columnwidth]{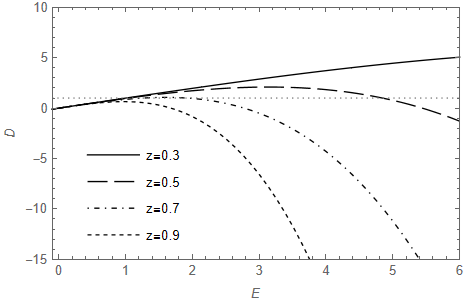}
\par
\end{centering}
\caption{Constitutive relation $D(E,z)$ for representative values of
the holographic coordinate. Near the AdS boundary the constitutive
response is essentially linear, whereas toward the infrared region the
quartic interaction progressively deforms the curve into a cubic
profile. The horizontal line $D=\rho$ intersects the constitutive
curve at a unique positive solution for $z<z_c$, becomes tangent at
$z=z_c$, and no longer intersects the physical branch for $z>z_c$, 
illustrating the loss of invertibility of the constitutive map}
\label{fig:constitutive}
\end{figure}

An alternative and particularly transparent interpretation of the same
critical behavior is provided by the constitutive curves shown in
Fig.~\ref{fig:constitutive}. Since the nonlinear correction is
proportional to $z^{4}$, increasing the holographic coordinate
progressively deforms the constitutive function from the linear
Maxwell relation into a non-monotonic cubic curve. The nonlinear
effects therefore become increasingly important as one moves toward
the infrared region of the bulk.

The extrema of the constitutive curve satisfy
\begin{equation}
\frac{\partial D}{\partial E}
=
1-3\gamma z^{4}E^{2}
=
0,
\label{eq:critical_condition}
\end{equation}
which is precisely the condition defining the vanishing of the
differential dielectric response. Consequently, the geometrical
tangency displayed in Fig.~\ref{fig:constitutive} is completely
equivalent to the analytical condition derived above.

For $z<z_c$, the constitutive relation remains monotonic over the
physical branch continuously connected to Maxwell electrodynamics, so
that the equation
\[
D(E,z)=\rho
\]
possesses a unique positive solution. At the critical point
$z=z_c$, the horizontal line $D=\rho$ becomes tangent to the
constitutive curve, signaling the loss of local invertibility.
Finally, for $z>z_c$, the maximum value of $D(E)$ becomes smaller than
the conserved displacement $\rho$, and the physical positive branch
ceases to exist. Although the cubic equation still admits a negative
solution, this branch cannot be continuously connected with the
boundary condition
\[
E(0)=\rho>0,
\]
and therefore does not describe the electrostatic background relevant
for the present holographic model.

For the physical branch to remain smooth throughout the full
holographic interval, the critical point must lie beyond the event
horizon,
\begin{equation}
z_c>1.
\end{equation}
The limiting case $z_c=1$ defines the endpoint of the regular branch.
Consequently,
\begin{equation}
\gamma\rho^2<\frac{4}{27},
\end{equation}
while equality determines the critical coupling.

Expressed in terms of the original higher-derivative coupling,
\begin{equation}
\alpha'
<
\sqrt{\frac{8}{81}}
\,
\frac{\beta}{\rho},
\label{eq:alpha_bound}
\end{equation}
establishing the condition for the existence of a globally regular
electrostatic solution. For the representative choice
$\rho=\beta=1$, one finds
\begin{equation}
\alpha'_{\rm crit}
=
\sqrt{\frac{8}{81}}
\simeq
0.314.
\end{equation}
This analytical prediction is fully consistent with the numerical
solutions of Fig.~\ref{fig:Radial-electric-field}, where regular
electrostatic configurations exist only for
$\alpha'<\alpha'_{\rm crit}$. The nonlinear constitutive relation therefore imposes an upper bound on the higher-derivative coupling if one requires a globally regular electrostatic background extending from the AdS boundary to the black-hole horizon.

The constitutive formulation developed in this subsection provides a
simple geometrical interpretation of the nonlinear gauge dynamics.
Rather than modifying the conservation of the radial electric flux,
the higher-derivative interaction changes the nonlinear mapping
between the conserved displacement field and the physical electric
field. This viewpoint not only explains the emergence of the critical
point and the upper bound on the nonlinear coupling, but also provides
the natural starting point for the perturbative expansion developed in
the following subsection.

\subsection{Chemical Potential and Thermodynamic Scale}

Although the constitutive relation, Eq.~(\ref{eq:constitutive}), is exact, explicit analytical expressions for 
the chemical potential still require an approximate expression for the electric field.
In the weak-coupling
regime, defined by $\gamma\rho^{2}\ll1$, the electric field may therefore
be expanded perturbatively as 
\begin{equation}
E(z)=E_{0}(z)+\gamma E_{1}(z)+\gamma^{2}E_{2}(z)+\mathcal{O}(\gamma^{3}).\label{eq:Eexpansion}
\end{equation}
Substituting Eq.~(\ref{eq:Eexpansion}) into Eq.~(\ref{eq:constitutive})
and collecting equal powers of $\gamma$ yields 
\begin{align}
E_{0} & =\rho,\nonumber \\
E_{1} & =\rho^{3}z^{4},\nonumber \\
E_{2} & =3\rho^{5}z^{8},
\end{align}
so that the electric field becomes 
\begin{equation}
E(z)=\rho+\gamma\rho^{3}z^{4}+3\gamma^{2}\rho^{5}z^{8}+\mathcal{O}(\gamma^{3}).\label{eq:Eperturbative}
\end{equation}

The chemical potential is determined by the asymptotic value of the
electrostatic potential, 
\begin{equation}
\mu=\phi(0)=\int^{1}_{0}E(z)\,dz.\label{eq:mu_integral}
\end{equation}
Substituting Eq.~(\ref{eq:Eperturbative}) immediately
gives 
\begin{equation}
\mu=\rho+\frac{\gamma\rho^{3}}{5}+\frac{\gamma^{2}\rho^{5}}{3}+\mathcal{O}(\gamma^{3}),\label{eq:mu_gamma}
\end{equation}
or, in terms of the original nonlinear coupling, 
\begin{equation}
\mu=\rho+\frac{3\alpha'^{2}\rho^{3}}{10\beta^{2}}+\frac{3\alpha'^{4}\rho^{5}}{4\beta^{4}}+\mathcal{O}(\alpha'^{6}).\label{eq:mu_alpha}
\end{equation}

Therefore, for fixed charge density, the higher-derivative interaction
increases the chemical potential relative to its Maxwell value. This
modification originates entirely from the nonlinear gauge dynamics
and does not require any backreaction on the gravitational background.

The Hawking temperature of the Reissner--Nordström geometry remains
\begin{equation}
T=\frac{\beta(3-q^{2})}{4\pi},\label{eq:T_repeat}
\end{equation}
which is independent of the nonlinear coupling within the probe approximation.
Nevertheless, the dimensionless thermodynamic scale governing the
dual field theory is modified through the chemical potential. Combining
Eqs.~(\ref{eq:mu_gamma}) and~(\ref{eq:T_repeat}) yield 
\begin{equation}
\frac{T}{\mu}=\frac{\beta(3-q^{2})}{4\pi\rho}\left[1+\frac{\gamma\rho^{2}}{5}+\frac{\gamma^{2}\rho^{4}}{3}+\mathcal{O}(\gamma^{3})\right]^{-1}.
\end{equation}
Expanding the inverse and using Eq.~(\ref{eq:gamma_definition}) for $\gamma$ gives 
\begin{equation}
\frac{T}{\mu}=\frac{\beta(3-q^{2})}{4\pi\rho}\left[1-\frac{3\alpha'^{2}\rho^{2}}{10\beta^{2}}-\frac{33\alpha'^{4}\rho^{4}}{50\beta^{4}}+\mathcal{O}(\alpha'^{6})\right].\label{eq:Tmu_alpha}
\end{equation}

The analytical results obtained in this section establish the complete analytical structure of the nonlinear gauge sector. The existence of an exact first integral, together with its interpretation as a nonlinear constitutive relation, reveals a critical point beyond which globally regular electrostatic solutions cease to exist. In the weak-coupling regime, explicit expressions for the chemical potential and for the thermodynamic ratio $T/\mu$ are obtained analytically. These results provide the analytical framework for investigating the fermionic response in the nonlinear electromagnetic background.

It is worth emphasizing that, within the present truncated probe description, the critical value of $\alpha'$ should be interpreted conservatively as the endpoint of the globally regular electrostatic branch rather than as direct evidence for a thermodynamic phase transition in the dual theory. What breaks down at that point is the smooth continuation of the physical electrostatic solution, due to the loss of invertibility of the constitutive relation, and not necessarily the existence of a more complete description once backreaction or additional higher-order operators are included. From this perspective, $\alpha'_{\mathrm{crit}}$ marks the boundary of validity of the regular branch captured by the present model.

\section{Fermionic Response in the Nonlinear Gauge Background}
\label{sec:fermionic_response}

Having established the analytical structure of the nonlinear gauge
sector, we now investigate how the corresponding electrostatic
background affects charged probe fermions. The fermionic degrees of
freedom are treated in the probe approximation, so that they propagate
on the fixed Reissner--Nordström-AdS geometry and couple minimally to
the nonlinear electrostatic potential obtained from the constitutive
relation derived in Sec.~\ref{model}I.

The central quantity of interest is the retarded Green's function of the dual fermionic operator, computed using the standard holographic prescription for fermions at finite density \cite{IqbalLiu2009Spinors,Lee2009,LiuMcGreevyVegh2011,FaulknerLiuMcGreevyVegh2011}. Since the nonlinear correction modifies
the radial profile of the electrostatic potential, it enters the
fermionic flow equations through the combination
\begin{equation}
\omega+q_f\phi(z),
\end{equation}
thereby changing the effective radial energy experienced by the bulk
fermion. Consequently, even though the Dirac equation itself retains the
same form as in the Maxwell case, the nonlinear gauge sector modifies
the boundary spectral response through the background function
$\phi(z)$.

\subsection{Numerical Setup}

The nonlinear electrostatic potential is obtained from the radial
electric field,
\begin{equation}
\phi(z)=\int_z^1 E(s)\,ds,
\label{eq:phi_from_E_fermions}
\end{equation}
where $E(z)$ is determined by the algebraic constitutive equation, Eq.~(\ref{eq:constitutive}).
Only the physical branch continuously connected to the Maxwell limit is
used. Therefore, the numerical analysis is restricted to the regime in
which a globally regular positive solution exists throughout the
interval $0\le z\le 1$.

For each background configuration, the fermionic flow equations given in Eq.~(\ref{eq:riccati_flow_reduced}) are
solved for the ratios defined in Eq.~(\ref{eq:xi_definition}).
The infalling boundary condition at the horizon is imposed as
\begin{equation}
\xi_I(1)=i,
\qquad I=1,2,
\label{eq:infalling_fermion}
\end{equation}
for nonzero frequency. In practice, the integration is started slightly
outside the horizon, at $z=1-\epsilon$, with $\epsilon\ll1$, and the
boundary condition is imposed by continuity from the infalling solution.

The retarded Green's function is extracted near the AdS boundary as
\begin{equation}
G_R(\omega,k)
=
\lim_{z\rightarrow0}
z^{-2m}
\begin{pmatrix}
\xi_1(z) & 0\\
0 & \xi_2(z)
\end{pmatrix}.
\label{eq:GR_numeric}
\end{equation}
The corresponding fermionic spectral function is defined by
\begin{equation}
A(\omega,k)
=
{\rm Im}\,{\rm Tr}\,G_R(\omega,k).
\label{eq:spectral_function}
\end{equation}
This quantity measures the density of fermionic excitations in the dual boundary theory and constitutes the primary diagnostic of the holographic fermionic response. In particular, the spectral function is the standard observable used to identify holographic Fermi surfaces through the appearance of quasiparticle-like peaks near the Fermi momentum \cite{Lee2009,CubrovicZaanenSchalm2009,LiuMcGreevyVegh2011,FaulknerLiuMcGreevyVegh2011}. In the following, it will be used to determine both the location of the Fermi momentum and the coherence of the corresponding fermionic excitation.

In the numerical analysis below, the Maxwell case $\alpha'=0$ is used
as the reference configuration. Finite values of $\alpha'$ are then
introduced within the regularity bound derived in Sec.~\ref{model}I. This allows
us to isolate the effect of the quartic electromagnetic interaction on
the fermionic spectral response while keeping the geometry fixed.

\subsection{Expected Effect of the Nonlinear Gauge Sector}

Before presenting the numerical spectra, it is useful to anticipate the
main qualitative effect of the nonlinear gauge interaction. For fixed
charge density, the perturbative result obtained in Sec.~\ref{model}I shows that
the nonlinear correction increases the chemical potential,
\begin{equation}
\mu
=
\rho
+
\frac{3\alpha'^2\rho^3}{10\beta^2}
+
\frac{3\alpha'^4\rho^5}{4\beta^4}
+
O(\alpha'^6).
\label{eq:mu_fermion_section}
\end{equation}
At fixed Hawking temperature, this implies a decrease of the
dimensionless ratio $T/\mu$,
\begin{equation}
\frac{T}{\mu}
=
\frac{\beta(3-q^2)}{4\pi\rho}
\left[
1
-
\frac{3\alpha'^2\rho^2}{10\beta^2}
-
\frac{33\alpha'^4\rho^4}{50\beta^4}
+
O(\alpha'^6)
\right].
\label{eq:Tmu_fermion_section}
\end{equation}

Since the fermionic spectral function is sensitive to the temperature-to-chemical-potential ratio, part of the sharpening of the spectral peak can already be anticipated from the decrease of $T/\mu$ even in the absence of gravitational backreaction. An enhancement of spectral weight therefore does not by itself imply a qualitatively new fermionic regime, but is naturally consistent with reduced effective thermal broadening.

The flow equation, however, depends on the full radial profile $\phi(z)$, not only on its boundary value $\mu$. This distinction is physically useful because not all spectral modifications have the same origin. An increase in the chemical potential at fixed Hawking temperature lowers the effective ratio $T/\mu$, thereby naturally reducing thermal broadening and contributing to the sharpening of the spectral peak. The nonlinear gauge interaction, however, also modifies the complete radial electrostatic profile entering the bulk Dirac equation. Consequently, the evolution of the fermionic spectrum reflects the combined action of these two effects: the reduction of the effective temperature scale, which favors sharper excitations, and the deformation of the bulk electrostatic background, which renormalizes quantities such as the Fermi momentum and the low-energy dispersion. The numerical results below are designed to disentangle these effects by comparing spectra obtained for different values of $\alpha'$, $q$, and $\beta$. This sensitivity to both the effective temperature scale and the full radial electrostatic potential is consistent with previous analyses of holographic fermion spectra in charged black-hole backgrounds \cite{Lee2009,LiuMcGreevyVegh2011,FaulknerLiuMcGreevyVegh2011}.

\subsection{Spectral Function and Fermi Momentum}

Having established the analytical properties of the nonlinear gauge background, we now investigate their impact on the fermionic response. The low-frequency spectral function is the primary observable characterizing the fermionic response. A holographic Fermi surface is identified by the emergence of a sharp quasiparticle-like peak near $\omega=0$, whose position defines the Fermi momentum. Numerically, the Fermi momentum is therefore determined from
\begin{equation}
k_F:\qquad
A(\omega\simeq0,k)
\ \ \text{is maximal}.
\label{eq:kF_definition}
\end{equation}

In the Maxwell limit ($\alpha'=0$), this prescription reproduces the standard holographic fermion spectrum. The quartic electromagnetic interaction continuously shifts the peak position and enhances its spectral weight, providing a direct observable signature of the nonlinear gauge dynamics in the dual strongly coupled theory.

\begin{figure}[h!]
\begin{centering}
\includegraphics[width=0.7\columnwidth]{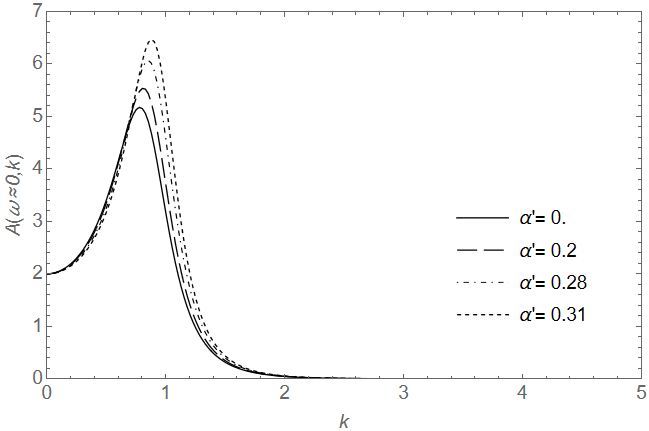}
\par\end{centering}
\caption{
Momentum dependence of the fermionic spectral function
$A(\omega\simeq0,k)$ for different values of the nonlinear coupling
$\alpha'$. Increasing the quartic gauge interaction shifts the
spectral peak toward larger momenta while simultaneously enhancing its
intensity, indicating a systematic modification of the holographic
Fermi surface.}
\label{fig:SpectralA}
\end{figure}

The momentum dependence of the spectral function at $\omega\simeq0$ is displayed in Fig.~\ref{fig:SpectralA} for representative values of the nonlinear coupling satisfying the regularity bound derived in Sec.~\ref{sec:analytical_gauge}.

Figure~\ref{fig:SpectralA} shows that the spectral function develops a well-defined quasiparticle-like peak for every value of $\alpha'$ considered. The persistence of this sharp maximum demonstrates that the holographic Fermi surface remains robust under the quartic electromagnetic correction. Nevertheless, both the peak position and its spectral weight evolve consistently as the nonlinear interaction is increased.

\begin{table}[h!]
\begin{centering}
\begin{tabular}{|c|c|c|c|c|}
\hline
$\alpha'$ & $\mu$ & $T/\mu$ & $k_F$ & $A_{\mathrm{max}}$ \\
\hline
\hline
0.00 & 1.00 & 0.238 & 0.779 & 5.17 \\
\hline
0.20 & 1.01 & 0.235 & 0.804 & 5.53 \\
\hline
0.28 & 1.03 & 0.232 & 0.854 & 6.07 \\
\hline
0.31 & 1.04 & 0.229 & 0.879 & 6.47 \\
\hline
\end{tabular}
\par
\end{centering}
\caption{Chemical potential, effective temperature ratio, Fermi momentum, and maximum spectral weight for representative values of the nonlinear coupling $\alpha'$.}
\label{tab:SpectralA}
\end{table}

The corresponding numerical values are summarized in Table~\ref{tab:SpectralA}.Two clear trends emerge from the numerical results. First, the Fermi momentum increases monotonically with the nonlinear coupling, shifting from
$k_F\simeq0.779$
for the Maxwell background to
$k_F\simeq0.879$
close to the critical coupling. Second, the maximum spectral weight also increases continuously, the spectral response becomes progressively sharper, consistent with enhanced coherence of the low-energy excitation as the nonlinear gauge interaction is strengthened.

\begin{figure}[h!]
\centering
\begin{minipage}{.5\textwidth}
  \centering
  \includegraphics[width=1.\linewidth]{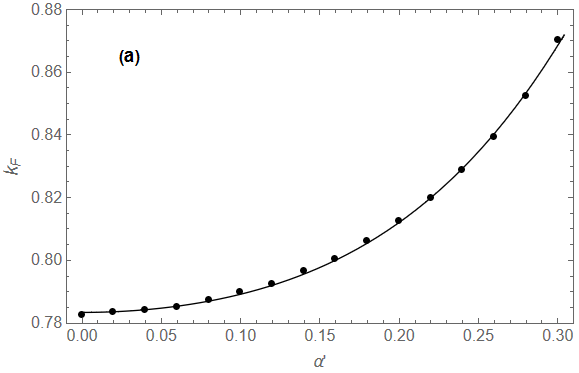}

\end{minipage}%
\begin{minipage}{.5\textwidth}
  \centering
  \includegraphics[width=1.\linewidth]{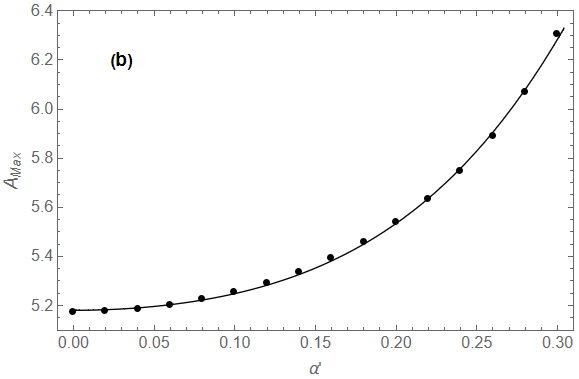}  
\end{minipage}
\caption{
Dependence of the fermionic observables on the nonlinear coupling
$\alpha'$. (a) Fermi momentum $k_F$ as a function of $\alpha'$. (b)
Maximum spectral weight
$A_{\mathrm{max}}=A(\omega\simeq0,k_F)$
as a function of $\alpha'$. Filled circles denote the numerical data,
while the solid curves represent least-squares polynomial fits
containing only even powers of $\alpha'$. The excellent agreement
confirms the even-power dependence predicted by the analytical
structure of the nonlinear gauge sector discussed in Sec.~\ref{sec:analytical_gauge}.
}
\label{fig:FermiMom}
\end{figure}

As an illustration of these trends, Fig.~\ref{fig:FermiMom} shows the dependence of both the Fermi momentum and the maximum spectral weight,
$A_{\mathrm{max}}=A(\omega\simeq0,k_F)$,
on the nonlinear coupling.

Motivated by the analytical structure of the gauge equations, we fitted the numerical data using polynomial expansions containing only even powers of the nonlinear coupling. The resulting least-squares fits are
\begin{align}
k_F(\alpha')
&=
0.783399
+
0.533812\,\alpha'^2
+
4.58103\,\alpha'^4,
\label{eq:kF_fit}
\\
A_{\mathrm{max}}(\alpha')
&=
5.18029
+
6.10297\,\alpha'^2
+
68.4916\,\alpha'^4.
\label{eq:Amax_fit}
\end{align}

The excellent agreement between the numerical data and the fitted curves demonstrates that both observables inherit the same even-power dependence on $\alpha'$ that characterizes the nonlinear gauge sector. This behavior is not merely empirical but follows directly from the analytical structure derived in Sec.~\ref{model}I, since the higher-derivative interaction enters the bulk equations exclusively through the effective parameter
\[
\gamma=\frac{3\alpha'^2}{2\beta^2}.
\]
Consequently, the fermionic observables investigated here naturally admit an expansion in even powers of $\alpha'$.

From Eq.~(\ref{eq:kF_fit}), the Fermi momentum increases by approximately $11\%$ over the interval
$0\le\alpha'\le0.30$, whereas Eq.~(\ref{eq:Amax_fit}) predicts an increase of approximately $22\%$ in the maximum spectral weight over the same range.

These numerical results are in excellent agreement with the analytical predictions obtained in Sec.~\ref{sec:analytical_gauge}. The quartic gauge interaction increases the chemical potential while leaving the Hawking temperature unchanged, thereby reducing the effective ratio $T/\mu$. The corresponding suppression of thermal broadening naturally sharpens the fermionic spectral peak. Within the present probe setup, this reduction of the effective temperature scale appears to provide the dominant explanation for the growth of $A_{\mathrm{max}}$. At the same time, the nonlinear correction modifies the complete radial electrostatic potential entering the Dirac equation through the combination
\[
\omega+q_f\phi(z),
\]
thereby changing the effective energy experienced by the bulk fermion and producing the observed displacement of the Fermi momentum. The evolution of the fermionic spectrum is therefore governed by two complementary mechanisms: the reduction of the effective temperature scale and the modification of the radial electrostatic background.

\section{Spectroscopic and Transport Signatures}
\label{sec:spectroscopy_transport}
The results obtained in the previous section demonstrate that the quartic
electromagnetic interaction modifies both the position of the holographic
Fermi surface and the spectral weight of the associated fermionic excitation.
To place these effects in a broader condensed-matter context, it is useful to
analyze the response of the system from two complementary perspectives.

From the holographic viewpoint, the observables considered in this section
correspond to different response functions of the same strongly coupled
system. The fermionic spectral function is naturally connected with
angle-resolved photoemission spectroscopy (ARPES), whereas the transverse
gauge perturbation probes the linear electromagnetic response underlying
the optical conductivity. Although the latter should not be identified
directly with the photoemission process itself, both observables originate
from the response of the same nonlinear holographic background to external
electromagnetic probes. They therefore provide complementary experimental
signatures of the underlying quartic gauge interaction.

The fermionic spectral function,
\begin{equation}
    A(\omega,k)=
    -\frac{1}{\pi}
    \operatorname{Im}\operatorname{Tr}G_{R}(\omega,k),
\end{equation}
characterizes the energy- and momentum-resolved single-particle response of
the boundary theory. It is therefore the holographic quantity most closely
related to the intrinsic spectral information underlying angle-resolved
photoemission spectroscopy. By contrast, the optical conductivity probes the
response of the conserved electric current to a spatially homogeneous,
time-dependent electric field and is determined by the retarded
current-current correlator,
\begin{equation}
    \sigma(\omega)
    =
    \frac{1}{i\omega}
    G^{R}_{J_xJ_x}(\omega,k=0).
\end{equation}
Although these quantities correspond to different experimental probes, they
are determined here within the same nonlinear gauge background. Computing
both responses therefore makes it possible to investigate whether the
quartic interaction produces correlated changes in single-particle
coherence and charge transport.

The holographic
fermionic Green's function describes a strongly coupled fermionic operator
of the boundary theory and is not automatically identical to the Green's
function of a microscopic electron in a specific material. A direct
quantitative comparison with ARPES intensities would require additional
information concerning photoemission matrix elements and, more generally,
a microscopic or semi-holographic embedding. Nevertheless, the evolution of
the peak position, dispersion, and linewidth of the holographic spectral
function provides a useful set of momentum-resolved observables that can be
compared with the phenomenology extracted from ARPES experiments.

\subsection{Momentum-resolved spectroscopic observables}
\label{subsec:spectroscopic_observables}

The analysis presented in Sec.~\ref{sec:fermionic_response} was based
primarily on the spectral function at very small frequency,
$A(\omega\simeq0,k)$, from which the Fermi momentum and maximum spectral
weight were extracted. A more detailed connection with momentum-resolved
spectroscopy requires considering the spectral function over a finite
frequency and momentum window around the Fermi surface.

Near a sufficiently well-defined fermionic excitation, the retarded Green's
function may be expressed schematically as
\begin{equation}
    G_R(\omega,k)
    \simeq
    \frac{Z}
    {
      \omega
      -v_F^{\ast}(k-k_F)
      -\Sigma(\omega,k)
    },
    \label{eq:quasiparticle_green}
\end{equation}
where $Z$ denotes the spectral residue, $v_F^{\ast}$ is the renormalized
Fermi velocity, and $\Sigma(\omega,k)$ is the fermionic self-energy. This
representation is used here only as a phenomenological parametrization of
the spectral peak and does not assume that the holographic state is a
Landau Fermi liquid.

The location of the Fermi momentum is determined, as before, by
\begin{equation}
    k_F:
    \qquad
    A(\omega\simeq0,k)
    \ \text{is maximal}.
\end{equation}
The local dispersion relation $\omega_{\star}(k)$ is then defined by the
position of the maximum of the spectral function at fixed momentum,
\begin{equation}
    A\bigl(\omega_{\star}(k),k\bigr)
    =
    \max_{\omega} A(\omega,k).
    \label{eq:peak_dispersion}
\end{equation}
In a sufficiently narrow region around the Fermi surface, the dispersion relation $\omega_{\star}(k)$ can be expanded as
\begin{equation}
\omega_{\star}(k)=\omega_0+v_F^{\ast}(k-k_F)+\cdots
\end{equation}
Therefore, the effective Fermi velocity may be estimated from
\begin{equation}
    v_F^{\ast}
    =
    \left.
    \frac{d\omega_{\star}(k)}{dk}
    \right|_{k=k_F}.
    \label{eq:effective_fermi_velocity}
\end{equation}

The coherence and lifetime of the excitation can be characterized by the
width of the spectral peak. Two complementary definitions are useful. At
fixed frequency, the momentum-distribution curve (MDC) is
\begin{equation}
    A_{\mathrm{MDC}}(k)
    =
    A(\omega_{\mathrm{fixed}},k),
\end{equation}
and its full width at half maximum (FWHM) will be denoted by
\begin{equation}
\Gamma_k(\omega)
=
\operatorname{FWHM}_{k}
\left[
A(\omega,k)
\right]_{\omega=\mathrm{const}},
\label{eq:mdc_width}
\end{equation}
Similarly, at fixed momentum the energy-distribution curve is
\begin{equation}
    A_{\mathrm{EDC}}(\omega)
    =
    A(\omega,k_{\mathrm{fixed}}),
\end{equation}
with frequency linewidth
\begin{equation}
\Gamma_{\omega}(k)
=
\operatorname{FWHM}_{\omega}
\left[
A(\omega,k)
\right]_{k=\mathrm{const}},
\label{eq:edc_width}
\end{equation}

In ARPES analyses, such linewidths are commonly used as measures of the
decay rate of single-particle excitations. Within the present holographic
model, their dependence on the nonlinear coupling provides a more robust
diagnostic of spectral coherence than the peak height alone. In particular,
the enhancement of $A_{\max}$ observed in Sec.~\ref{sec:fermionic_response}
should be accompanied by a reduction of $\Gamma_k$ and/or
$\Gamma_{\omega}$ if the quartic interaction genuinely produces a sharper
fermionic excitation rather than merely changing the overall normalization
of the Green's function.

\begin{figure}[h!]
\centering
 \includegraphics[width=1.01\linewidth]{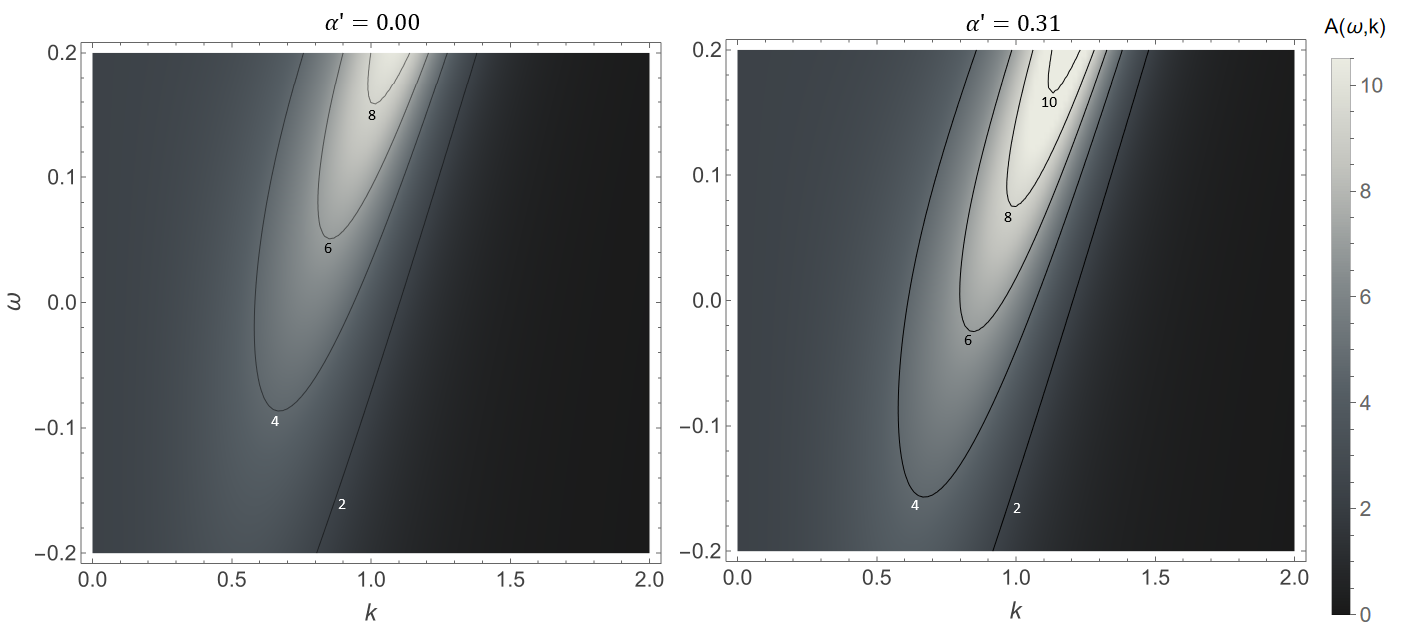}
\caption{Momentum-resolved fermionic spectral function
$A(\omega,k)$ for (a) the Maxwell background, $\alpha'=0$, and
(b) a nonlinear background close to the endpoint of the regular branch,
$\alpha'=0.31$. Both panels use the same frequency and momentum ranges
and a common color scale. Increasing the nonlinear coupling shifts the
spectral ridge toward larger momenta and concentrates spectral weight
around the dominant excitation.}
\label{fig:spectral}
\end{figure}

Figure~\ref{fig:spectral} displays the momentum-resolved spectral function
$A(\omega,k)$ for the Maxwell theory and for a representative value
of the quartic nonlinear coupling close to the endpoint of the regular
branch. Since the color scale represents the spectral weight, the bright
ridge directly identifies the dominant fermionic excitation in the
frequency--momentum plane.

Several features become immediately apparent. First, the ridge is
systematically displaced toward larger momenta as the nonlinear coupling
is increased, providing a direct visualization of the increase of the
Fermi momentum already inferred from the low-frequency spectra discussed
in \ref{sec:fermionic_response}. Second, the spectral weight becomes increasingly concentrated in the
vicinity of the dominant ridge. This behavior is consistent with the
enhancement of the peak height found in \ref{sec:fermionic_response}. A quantitative
assessment of the corresponding narrowing requires the MDC and EDC
linewidths discussed below.

The momentum-resolved maps also make it possible to extract
one-dimensional cuts commonly employed in ARPES analyses.
Horizontal cuts at fixed frequency define the momentum-distribution
curves (MDCs), whereas vertical cuts at fixed momentum define the
energy-distribution curves (EDCs). Their full widths at half maximum
provide independent estimates of the momentum and energy linewidths,
$\Gamma_k$ and $\Gamma_\omega$, respectively. 
For each value of the nonlinear coupling, the MDC linewidth was
extracted from the cut at $\omega \simeq 0$, whereas the EDC linewidth
was obtained from the cut at the corresponding Fermi momentum,
$k=k_F(\alpha')$. The linewidths were determined from Lorentzian fits to the dominant
MDC and EDC peaks, with $\Gamma_k$ and $\Gamma_\omega$ identified with
the corresponding full widths at half maximum.
The obtained values are reported in Table \ref{tab:SpectralB}.

The ridge also defines the quasiparticle dispersion
$\omega^\star(k)$, obtained by locating the maximum of the spectral
function at fixed momentum. In the vicinity of the Fermi surface the
dispersion remains approximately linear, allowing the renormalized
Fermi velocity to be extracted from its local slope.

\begin{figure}[h!]
\centering
 \includegraphics[width=0.8\linewidth]{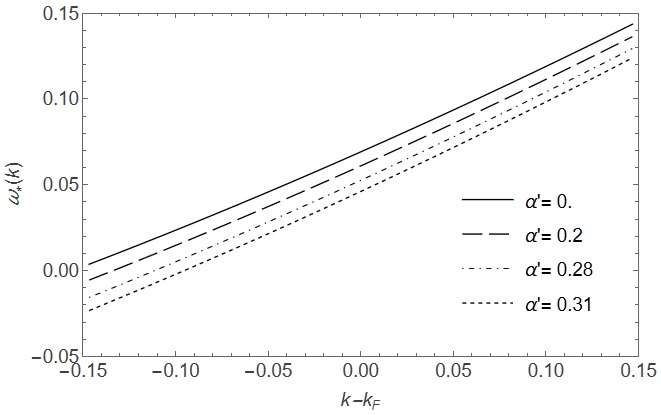}
\caption{Low-energy quasiparticle dispersion $\omega^\star(k)$ plotted
against the momentum measured relative to the corresponding Fermi
momentum, $k-k_F$, for representative values of the nonlinear coupling
$\alpha'$. The slope at $k-k_F=0$ determines the renormalized Fermi
velocity $v_F^\ast$.}
\label{fig:dispersion}
\end{figure}

The dispersions extracted from the spectral ridges are shown in
Fig.~\ref{fig:dispersion} as functions of the relative momentum
$k-k_F$, where the Fermi momentum is determined separately for each
value of $\alpha'$. This representation removes the overall
displacement of the Fermi surface and isolates the change in the local
slope of the dispersion. Within the frequency window considered, the
dispersions remain approximately linear around $k-k_F=0$. Their slopes
therefore provide direct estimates of the renormalized Fermi velocity,
which increases moderately from
$v_F^\ast \simeq 0.477$ in the Maxwell case to
$v_F^\ast \simeq 0.502$ near the endpoint of the regular branch, as
summarized in Table~\ref{tab:SpectralB}. The persistence of an approximately linear low-energy dispersion
demonstrates that the quartic nonlinear interaction continuously
renormalizes the effective quasiparticle parameters---such as
$k_F$, $v_F^\ast$, and the spectral weight---without destroying the
underlying holographic Fermi surface.

\begin{table}[h!]
\begin{centering}
\begin{tabular}{|c|c|c|c|}
\hline
$\alpha'$ & $v_{F}^{*}$ & $\Gamma_k(\omega)$ & $\Gamma_{\omega}(k)$  \\
\hline
\hline
0.00 & 0.477 & 0.696 & 0.292  \\
\hline
0.20 & 0.484 & 0.664 & 0.280  \\
\hline
0.28 & 0.495 & 0.622 & 0.265  \\
\hline
0.31 & 0.502 & 0.595 & 0.256  \\
\hline
\end{tabular}
\par
\end{centering}
\caption{Renormalized Fermi velocity and momentum and energy linewidths
extracted from the full spectral maps for representative values of the
nonlinear coupling. The velocity is obtained from a linear fit to the
local dispersion near $k_F$, while $\Gamma_k$ and $\Gamma_\omega$ are
the FWHM values extracted from Lorentzian fits to the corresponding MDC
and EDC profiles.}
\label{tab:SpectralB}
\end{table}

At the same time, both linewidths decrease monotonically with the
nonlinear coupling. The momentum linewidth changes from
$\Gamma_k\simeq0.696$ to $\Gamma_k\simeq0.595$, whereas the energy
linewidth decreases from $\Gamma_\omega\simeq0.292$ to
$\Gamma_\omega\simeq0.256$. These results confirm that the enhancement
of $A_{\max}$ found in Sec.~\ref{sec:fermionic_response} is accompanied
by a genuine narrowing of the spectral excitation, rather than merely
by an overall change in the normalization of the Green's function.

Taken together, the increase of $A_{\max}$, the moderate enhancement of
$v_F^\ast$, and the reduction of both momentum and energy linewidths
show that the quartic interaction modifies not only the location of the
Fermi surface but also the coherence and low-energy dispersion of the
fermionic excitation. In the following subsection, these
single-particle trends will be compared with the optical and DC current
responses, which probe a distinct collective channel of the boundary
theory.

\subsection{Optical conductivity}
\label{subsec:optical_conductivity}

To complement the single-particle spectral analysis, we now investigate
the linear response of the conserved boundary current. It is important
to distinguish the radial electric field,
\begin{equation}
E(z)=-\phi'(z),
\end{equation}
which determines the nonlinear electrostatic background through the bulk
field-strength component $F_{zt}$, from the electric field used to probe
the conductivity of the boundary theory. The latter is generated by the
time-dependent transverse perturbation introduced in Sec.~\ref{model},
\begin{equation}
\delta A_x(t,z)
=
a_x(z)e^{-i\omega t},
\qquad
\delta A_z=0,
\label{eq:transverse_gauge_perturbation}
\end{equation}
and corresponds to the field-strength component $F_{tx}$.

The linearized equation governing this perturbation was derived in
Eq.~(\ref{eq:A_nonlinear_reduced}). An important consequence is that the
same nonlinear response function $\mathcal K(z)$ appearing in the
electrostatic background also controls the propagation of the transverse
gauge fluctuation. Therefore, the nonlinear electrodynamic correction
simultaneously affects the equilibrium configuration, the fermionic
spectral function, and the optical conductivity.

The boundary electric field follows directly from the definition of the
field-strength tensor,
\begin{equation}
F_{tx}
=
\partial_tA_x-\partial_xA_t.
\end{equation}
Since the electrostatic background depends only on the holographic
coordinate,
$\partial_xA_t=0$,
one obtains
\begin{equation}
E_x
=
-F_{tx}
=
i\omega a_x(z)e^{-i\omega t}. 
\label{eq:axz}
\end{equation}
Near the AdS boundary,
\begin{equation}
a_x(z)
=
a_x^{(0)}
+
a_x^{(1)}z+\cdots,
\end{equation}
so that the amplitude of the externally applied electric field is simply
\begin{equation}
E_x^{(\mathrm{bdry})}
=
i\omega a_x^{(0)}.
\end{equation}

The induced current is determined holographically from the canonical
momentum conjugate to the transverse gauge field,
\begin{equation}
\Pi_x(z)
=
-\beta f(z)\mathcal K(z)a_x'(z).
\end{equation}
According to the holographic dictionary,
\begin{equation}
J_x=-\Pi_x(0).
\end{equation}
Since
$f(0)=1$,
$\mathcal K(0)=1$,
and
$a_x'(0)=a_x^{(1)}$,
the boundary current becomes
\begin{equation}
J_x
=
\beta a_x^{(1)}.
\end{equation}

Therefore, the leading coefficient of the near-boundary expansion acts
as the source for the external vector potential, whereas the subleading
coefficient determines the induced current. The optical conductivity is
then obtained from Ohm's law,
\begin{equation}
\sigma(\omega)
=
\frac{J_x}{E_x}
=
\frac{\beta a_x^{(1)}}
{i\omega a_x^{(0)}}.
\label{eq:optical_conductivity}
\end{equation}

In general, the optical conductivity is a complex response function,
\begin{equation}
\sigma(\omega)
=
\sigma_{1}(\omega)
+
i\,\sigma_{2}(\omega),
\end{equation}
where the real part,
$\sigma_{1}(\omega)=\mathrm{Re}\,\sigma(\omega)$,
describes the dissipative response associated with energy absorption by
the system, while the imaginary part,
$\sigma_{2}(\omega)=\mathrm{Im}\,\sigma(\omega)$,
characterizes the reactive response, corresponding to the temporary
storage and release of electromagnetic energy. Since the ingoing
boundary condition renders the near-boundary coefficients
$a_x^{(0)}$ and $a_x^{(1)}$ complex, both contributions arise
naturally from Eq.~(\ref{eq:optical_conductivity}) and encode
complementary information about the dynamical response of the dual
field theory.

\begin{figure}[h!]
\centering
\includegraphics[width=1.1\linewidth]{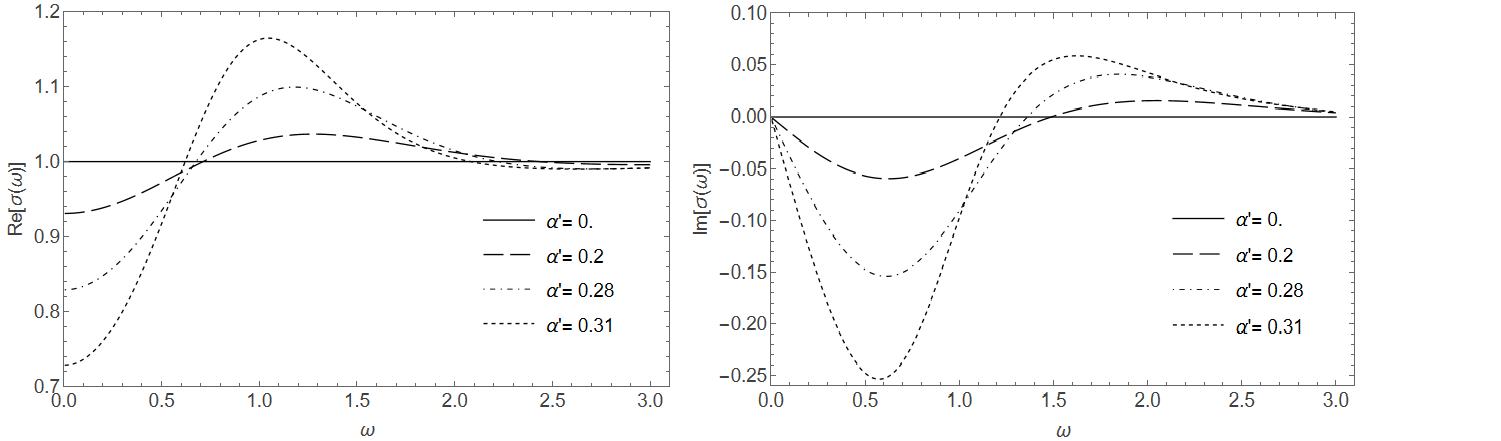}
\caption{Optical conductivity for representative values of the quartic
nonlinear coupling $\alpha'$. The left panel shows the dissipative
response, $\mathrm{Re}\,\sigma(\omega)$, and the right panel the
reactive response, $\mathrm{Im}\,\sigma(\omega)$. Increasing the
nonlinear coupling redistributes the optical response from low to
intermediate frequencies while simultaneously modifying the reactive
component. Together with the momentum-resolved spectroscopic
observables discussed above, these results demonstrate that the same
quartic gauge interaction produces correlated changes in both the
single-particle spectrum and the collective transport response of the
dual strongly coupled system.}
\label{fig:conductiv}
\end{figure}

In the following, we compute numerically both the real and imaginary
parts of the optical conductivity for different values of the nonlinear
coupling $\alpha'$, allowing the influence of quartic nonlinear
electrodynamics on the dissipative and reactive transport properties of
the boundary system to be analyzed separately. The results are shown in Fig.~\ref{fig:conductiv}.

The finite-frequency response reveals that the quartic nonlinear interaction does not simply suppress the conductivity over the whole spectrum. Instead, it redistributes spectral weight from the infrared region toward intermediate frequencies, producing a broad maximum whose amplitude increases as the endpoint of the regular branch is approached. This behavior is consistent with the redistribution of charge transport across different frequency scales and foreshadows the suppression of the DC conductivity discussed in the next subsection, while simultaneously modifying the reactive response. For larger frequencies, however, the optical conductivity gradually
approaches the Maxwell result, indicating that the influence of the
quartic nonlinear interaction becomes progressively weaker outside the
low-energy regime.

The imaginary part develops a pronounced negative minimum followed by a sign reversal at intermediate frequencies before gradually approaching the Maxwell behavior at higher frequencies. The systematic evolution of this structure accompanies the redistribution of spectral weight observed in the dissipative component and reflects the increasing influence of the nonlinear gauge interaction on the low-energy transport response.

The numerical results further suggest that the optical conductivity becomes progressively insensitive to the quartic nonlinear coupling at sufficiently large frequencies, with all curves tending toward the Maxwell response

\subsubsection{DC conductivity}
\label{subsubsec:dc_conductivity}

An immediate analytical result follows from the zero-frequency limit of
the fluctuation equation (\ref{eq:A_nonlinear_reduced}). In the limit
$\omega\rightarrow0$, one finds
\begin{equation}
\frac{d}{dz}
\left[
f(z)\mathcal K(z)a_x'(z)
\right]
=
0,
\end{equation}
which, according to the definition of the canonical momentum introduced
above, is equivalent to
\begin{equation}
\partial_z\Pi_x(z)=0.
\end{equation}

Therefore, the radial current is conserved along the holographic
direction, implying that it may be evaluated at any radial position.
The most convenient choice is the black-hole horizon, where the
infalling boundary condition completely fixes the near-horizon behavior
of the perturbation.

The retarded Green's function is obtained by imposing the standard
ingoing-wave boundary condition at the black-hole horizon
\cite{SonStarinets2002,KovtunStarinets2005},
\begin{equation}
a_x(z)
\sim
(1-z)^{-i\omega/(4\pi T)},
\qquad
z\rightarrow1,
\label{eq:ax_infalling}
\end{equation}
which ensures that the perturbation carries flux only into the horizon
and therefore yields the causal (retarded) response of the dual field
theory. 

Using expression (\ref{eq:ax_infalling}) together with
\begin{equation}
f(z)
\simeq
4\pi T(1-z),
\end{equation}
one obtains
\begin{equation}
\Pi_x(1)
=
-i\beta\omega
\mathcal K(1)
a_x(1).
\end{equation}

Since the electric field at the horizon is
\begin{equation}
E_x(1)
=
i\omega a_x(1),
\end{equation}
the DC conductivity becomes
\begin{equation}
\sigma_{DC}
=
\lim_{\omega\rightarrow0}
\frac{\Pi_x(1)}
{-E_x(1)}
=
\mathcal K(1).
\label{eq:sigma_dc_horizon}
\end{equation}

This remarkably simple expression for the DC conductivity shows that the
zero-frequency transport coefficient is completely determined by the
nonlinear electromagnetic response evaluated at the black-hole horizon.
Consequently, the influence of the quartic interaction on charge
transport can be understood directly from the effective response
function introduced in Sec.~\ref{model}I, without requiring the numerical
solution of the frequency-dependent fluctuation equation.

Using the nonlinear response function
$\mathcal K(z)$ introduced in Eq.~(\ref{nonlinearResponse}), 
Eq.~(\ref{eq:sigma_dc_horizon}) may be rewritten as
\begin{equation}
\sigma_{DC}=1-\gamma E(1)^2 = \frac{\rho}{E(1)}.
    \label{eq:sigma_dc_exact}
\end{equation}

This result establishes a direct relation between the boundary current
response and the nonlinear electric field at the horizon. In the Maxwell
limit ($\alpha'=0$), $\gamma\rightarrow0$ and $E\rightarrow\rho$, so that
\begin{equation}
    \sigma_{\mathrm{DC}}^{(\alpha'=0)}=1
\end{equation}
in the normalization adopted here.

The analytical dependence of the DC conductivity on the nonlinear
coupling follows immediately from the perturbative expansion of the
electric field obtained in Sec.~\ref{model}, Eq.~(\ref{eq:Eperturbative}), resulting in
\begin{equation}
    \sigma_{\mathrm{DC}}
    =
    1-\gamma\rho^2
    -2\gamma^2\rho^4
    +\mathcal{O}(\gamma^3).
    \label{eq:sigma_dc_gamma_expansion}
\end{equation}
or in terms of the original nonlinear coupling,
\begin{equation}
    \sigma_{\mathrm{DC}}
    =
    1
    -
    \frac{3\alpha'^2\rho^2}{2\beta^2}
    -
    \frac{9\alpha'^4\rho^4}{2\beta^4}
    +
    \mathcal{O}(\alpha'^6).
    \label{eq:sigma_dc_alpha_expansion}
\end{equation}

Using Eq~(\ref{eq:Ec}) for $E_c$,  the endpoint of the globally regular electrostatic branch, expression (\ref{eq:sigma_dc_exact}) becomes
\begin{equation}
  \sigma_{\mathrm{DC}}^{(\alpha'_{\mathrm{crit}})} 
    =
    \frac{2}{3}.
    \label{eq:sigma_dc_critical}
\end{equation}
Thus, within the normalization adopted here, the Maxwell theory
provides the largest DC conductivity, whereas the quartic interaction
continuously suppresses charge transport as the endpoint of the
globally regular electrostatic branch is approached.

Unlike the finite-frequency optical conductivity, whose complete
frequency dependence requires solving the fluctuation equation
numerically, the DC conductivity is obtained entirely analytically.
This result establishes a direct connection between the constitutive
structure of the nonlinear gauge sector and a measurable transport
coefficient of the boundary theory. It therefore complements the
momentum-resolved spectroscopic observables discussed above by showing
that the same quartic interaction which renormalizes the fermionic
spectrum also produces a systematic suppression of charge transport.

\subsection{Complementarity between single-particle and transport responses}
\label{subsec:complementarity}

The fermionic spectral function and the optical conductivity probe distinct
correlation functions of the boundary theory. The former is determined by a
fermionic two-point function and characterizes the propagation and decay of
single-particle excitations, whereas the latter is determined by a
current-current correlator and characterizes the relaxation of the electric
current. Consequently, a direct identification between the ARPES linewidth
and an optical transport rate is neither assumed nor generally expected.

Nevertheless, both responses are computed in the present model using the
same nonlinear electrostatic background. The quartic interaction changes the
radial profile $\phi(z)$ entering the Dirac equation and simultaneously
changes the effective coupling $\mathcal{K}(z)$ governing transverse gauge
fluctuations. This common origin makes it possible to investigate whether
changes in momentum-resolved spectral coherence are accompanied by
correlated changes in charge transport.

A particularly useful comparison is provided by the evolution of the
fermionic linewidths and the low-frequency optical response. If the increase
of $A_{\max}$ observed at larger $\alpha'$ is associated with a genuinely
longer-lived fermionic excitation, one expects a corresponding reduction of
$\Gamma_k$ and/or $\Gamma_{\omega}$. The optical conductivity can then be
used to determine whether this enhanced single-particle coherence is
accompanied by a more pronounced low-frequency current response.

Such a correspondence need not be one-to-one. Single-particle decay and
current relaxation weight scattering processes differently, and a sharper
fermionic spectral peak does not necessarily imply a larger DC
conductivity. A possible difference between the two trends would therefore
not constitute an inconsistency, but rather evidence that the quartic
interaction affects single-particle and collective response channels in
different ways.

The comparison with experiment should consequently be understood at the
level of qualitative spectral and transport trends. The holographic results
may be confronted with ARPES phenomenology through the evolution of
$k_F$, the quasiparticle dispersion, and the EDC and MDC linewidths, while
the optical conductivity supplies an independent probe of the collective
current response. Together, these observables provide a broader
characterization of how nonlinear bulk gauge dynamics is encoded in the
boundary theory.

\section{Conclusions}
\label{sec:conclusions}

In this work we investigated how the leading string-inspired quartic
correction $\alpha'^2F^4$ modifies both fermionic spectral properties
and charge transport in a bottom-up holographic model. Working within
the probe approximation, we focused on isolating the imprint of the
quartic nonlinear gauge interaction on fermionic spectroscopy and
transport, without mixing these effects with gravitational
backreaction. Within this controlled setting, a central result of the
present analysis is that the leading quartic interaction retains a
remarkable degree of analytical tractability despite the higher-order
correction. We derived an exact first integral of the nonlinear gauge
equation, reformulated the electrostatic problem as a nonlinear
constitutive relation, identified the endpoint of the globally regular
electrostatic branch, and obtained perturbative expressions for the
electric field, the chemical potential, and the thermodynamic ratio
$T/\mu$. These results provide a unified analytical framework for the
subsequent numerical analysis, while also yielding a transparent
physical interpretation of the nonlinear gauge dynamics.

Using these analytical backgrounds, we computed the retarded fermionic
Green's function and investigated the corresponding momentum-resolved
spectral observables. The quartic interaction preserves the
holographic Fermi surface while continuously renormalizing its
properties. Besides shifting the Fermi momentum and enhancing the
spectral weight, the nonlinear interaction produces a moderate
increase in the renormalized Fermi velocity together with systematic
reductions of both momentum and energy linewidths. Within the present
framework, these results indicate that the higher-order gauge
interaction tends to increase the coherence of the low-energy
quasiparticle excitations. Moreover, the observed dependence of the
fermionic observables on even powers of $\alpha'$ follows directly
from the analytical structure of the nonlinear gauge equations,
establishing a quantitative link between bulk nonlinear
electrodynamics and boundary spectral properties.

The extension of the analysis to optical transport shows that the same
nonlinear interaction also leaves clear signatures in the collective
response of the boundary theory. The optical conductivity exhibits a
redistribution of spectral weight from the infrared toward
intermediate frequencies, while gradually recovering the Maxwell
behavior at higher frequencies. At the same time, the DC conductivity
admits the simple analytical expression
$\sigma_{\mathrm{DC}}=\mathcal K(1)=\rho/E(1)$, showing that the
zero-frequency transport coefficient is fully determined by the
nonlinear electromagnetic response evaluated at the black-hole
horizon. Within the present normalization, the DC conductivity
decreases monotonically with increasing nonlinear coupling, reaching
$\sigma_{\mathrm{DC}}=2/3$ at the endpoint of the globally regular
electrostatic branch.

Taken together, these results show that the same nonlinear response
function $\mathcal K(z)$ governs the equilibrium electrostatic
background, the propagation of gauge-field fluctuations, the
fermionic spectral response, and the transport coefficients of the
dual theory. The present work therefore provides a unified picture of
how a leading quartic bulk nonlinearity propagates from the gauge
sector to both single-particle spectroscopic observables and
collective charge transport. In particular, the momentum-resolved
quantities analyzed here strengthen the connection between
holographic fermion models and experimentally accessible observables
probed by ARPES, while the optical and DC conductivities provide
complementary information on the collective response.

It is important, however, to state clearly the domain of
interpretation of the present results. Since the analysis is
performed in the probe limit and within a truncation to the leading
quartic correction, the conclusions should be understood as
identifying the qualitative trends and constitutive structure
generated by the leading $F^4$ interaction, rather than as a complete
description of the fully backreacted bulk theory or of the full
string effective action. In particular, the critical coupling
identified here should be interpreted conservatively as the endpoint
of the globally regular branch of the truncated quartic effective
theory, associated with the loss of invertibility of the constitutive
relation, rather than as direct evidence for a thermodynamic phase
transition or as a universal bound of the complete theory.

Several natural extensions of the present analysis deserve future
investigation, including the incorporation of gravitational
backreaction, the study of additional higher-order corrections in the
effective action, the inclusion of magnetic fields and
magnetotransport, and more systematic comparisons with other nonlinear
gauge sectors such as Born--Infeld and power-Maxwell
electrodynamics. Such developments should help clarify which of the
spectroscopic and transport signatures identified here are specific to
the quartic constitutive structure and which instead represent more
general consequences of nonlinear holographic electrodynamics.

\acknowledgments
Vitor Neves da Silva acknowledges CAPES (Grant No. 88887.846973/2023-00)
for financial support.

% Bibliography

%% [A] Recommended: using JHEP.bst file
 \bibliographystyle{JHEP}
\bibliography{biblio.bib}

%% or
%% [B] Manual formatting (see below)
%% (i) We suggest to always provide author, title and journal data or doi:
%% in short all the informations that clearly identify a document.
%% (ii) please avoid comments such as "For a review'', "For some examples",
%% "and references therein" or move them in the text. In general, please leave only references in the bibliography and move all
%% accessory text in footnotes.
%% (iii) Also, please have only one work for each \bibitem.

%%\begin{thebibliography}{99}

\

%%\end{thebibliography}
\end{document}